\documentclass[12pt]{article}
\usepackage{amssymb}
\def\be{\begin{equation}}
\def\ee{\end{equation}}
\def\bea{\begin{eqnarray}}
\def\eea{\end{eqnarray}}
\def\p{\partial}

\def\half{{\textstyle{1\over2}}}
\def\halfi{{\textstyle{i\over2}}}
\def\three{{\textstyle{1\over3}}}

\def\np#1{{\sl Nucl.~Phys.~\bf B#1}}
\def\pl#1{{\sl Phys.~Lett.~\bf B#1}}
\def\pr#1{{\sl Phys.~Rev.~\bf D#1}}
\def\prl#1{{\sl Phys.~Rev. Lett.~\bf #1}}

\catcode`\@=11
\def\@citex[#1]#2{%
\if@filesw \immediate \write \@auxout {\string \citation {#2}}\fi
\@tempcntb\m@ne \let\@h@ld\relax \def\@citea{}%
\@cite{%
  \@for \@citeb:=#2\do {%
    \@ifundefined {b@\@citeb}%
      {\@h@ld\@citea\@tempcntb\m@ne{\bf ?}%
      \@warning {Citation `\@citeb ' on page \thepage \space undefined}}%
      {\@tempcnta\@tempcntb \advance\@tempcnta\@ne%
      \@tempcntb\number\csname b@\@citeb \endcsname \relax%
      \ifnum\@tempcnta=\@tempcntb 
        \ifx\@h@ld\relax%
          \edef \@h@ld{\@citea\csname b@\@citeb\endcsname}%
        \else%
          \edef\@h@ld{\ifmmode{-}\else--\fi\csname b@\@citeb\endcsname}%
        \fi%
      \else
        \@h@ld\@citea\csname b@\@citeb \endcsname%
        \let\@h@ld\relax%
      \fi}%
    \def\@citea{,\penalty\@highpenalty\,}%
  }\@h@ld
}{#1}}

\def\@citeb#1#2{{[#1]\if@tempswa , #2\fi}}
\def\@citeu#1#2{{$^{#1}$\if@tempswa , #2\fi }}
\def\@citep#1#2{{#1\if@tempswa , #2\fi}}

\def\bcites{         
        \catcode`\@=11
        \let\@cite=\@citeb
        \catcode`\@=12
}

\def\upcites{         
        \catcode`\@=11
        \let\@cite=\@citeu
        \catcode`\@=12
}

\def\plaincites{      
        \catcode`\@=11
        \let\@cite=\@citep
        \catcode`\@=12
}

\newcount\hour
\newcount\minute
\newtoks\amorpm
\hour=\time\divide\hour by 60
\minute=\time{\multiply\hour by 60 \global\advance\minute by-\hour}
\edef\standardtime{{\ifnum\hour<12 \global\amorpm={am}%
        \else\global\amorpm={pm}\advance\hour by-12 \fi
        \ifnum\hour=0 \hour=12 \fi
        \number\hour:\ifnum\minute<10 0\fi\number\minute\the\amorpm}}
\edef\militarytime{\number\hour:\ifnum\minute<10 0\fi\number\minute}

\def\draftlabel#1{{\@bsphack\if@filesw {\let\thepage\relax
   \xdef\@gtempa{\write\@auxout{\string
      \newlabel{#1}{{\@currentlabel}{\thepage}}}}}\@gtempa
   \if@nobreak \ifvmode\nobreak\fi\fi\fi\@esphack}
        \gdef\@eqnlabel{#1}}
\def\@eqnlabel{}
\def\@vacuum{}
\def\marginnote#1{}
\def\draftmarginnote#1{\marginpar{\raggedright\scriptsize\tt#1}}
\def\draft{
        \pagestyle{plain}
        \overfullrule=2pt
        \oddsidemargin -.5truein
        \def\@oddhead{\sl \phantom{\today\quad\militarytime} \hfil
        \smash{\Large\sl DRAFT} \hfil \today\quad\militarytime}
        \let\@evenhead\@oddhead
        \let\label=\draftlabel
        \let\marginnote=\draftmarginnote
        \def\ps@empty{\let\@mkboth\@gobbletwo
        \def\@oddfoot{\hfil \smash{\Large\sl DRAFT} \hfil}
        \let\@evenfoot\@oddhead}
        \def\@eqnnum{(\theequation)\rlap{\kern\marginparsep\tt\@eqnlabel}%
        \global\let\@eqnlabel\@vacuum}  }

\begin{document}

\hfill UTHET-04-0801

\vspace{-0.2cm}

\begin{center}
\Large
{\bf Unitarity, quasi-normal modes and the AdS$_3$/CFT$_2$ correspondence}
\normalsize

\vspace{0.8cm}
{\sc George Siopsis}
\footnote{
Research supported by the US Department of Energy under grant
DE-FG05-91ER40627.}
\\ {\em Department of Physics
and Astronomy, \\
The University of Tennessee, Knoxville, \\
TN 37996 - 1200, USA.} \\
{\tt email: gsiopsis@utk.edu}

\end{center}

\vspace{0.8cm}
\large
\centerline{\bf Abstract}
\normalsize
\vspace{.5cm}

In general, black-hole perturbations are governed by a discrete spectrum of
complex eigen-frequencies (quasi-normal modes). This signals the breakdown of
unitarity. In asymptotically AdS spaces, this is puzzling because the corresponding CFT is unitary.
To address this issue in three dimensions, we replace the BTZ black hole by a
wormhole, following a suggestion by Solodukhin [hep-th/0406130]. We solve the
wave equation for a massive scalar field and find an equation for the poles of the propagator.
This equation yields a rich spectrum of {\em real} eigen-frequencies. We
show that the throat of the wormhole is $o(e^{-1/G})$, where $G$ is Newton's
constant. Thus, the quantum effects which might produce the wormhole are
non-perturbative. 

\newpage

Quasi-normal modes govern the time evolution of a perturbed black hole. Typically, they form a discrete spectrum of {\em complex} frequencies
which are derived as the eigenvalues of a wave equation in the black-hole
background.
Their imaginary part is negative implying that the black hole eventually relaxes
back to its original (thermal) equilibrium at (Hawking) temperature $T_H$~\cite{bibq2}.
This is due to leakage of information into the horizon and signals the breakdown
of unitarity. It is closely related to Hawking's information loss paradox~\cite{bibH}.
This issue is still unresolved~\cite{bibbh1,bibbh2,bibbh3}.
Its resolution will require an understanding of quantum gravity beyond its
semi-classical approximation.

In asymptotically AdS space-times, there is an additional tool in tackling the
problem of non-unitarity, due to the AdS/CFT correspondence~\cite{bibadscft}. The complex
quasi-normal frequencies are poles of the retarded propagator in the
corresponding CFT living on the boundary of AdS space. This is puzzling, because
the CFT is unitary and therefore the propagator should possess {\em real} poles
only. On the gravity side, the Poincar\'e recurrence theorem implies that, e.g.,
a two-point function would be quasi-periodic with a period $t_P$, say. For times
$t\ll t_P$, the system may look like it is decaying back to thermal equilibrium,
but for $t\gtrsim t_P$, it should return to its original state (or close) an
infinite number of times. In fact, the
theorem guarantees that the system will {\em never} relax back to its original
state.

This problem has been tackled in three dimensions, where exact results can be
derived~\cite{bibq7,bibq12,bibq13}. AdS$_3$ arises in type IIB superstring theory in the near horizon limit
of a large number of D1 and D5 branes~\cite{bibadscft}. Low energy excitations form a gas of
strings wound around a circle with winding number $\mathtt{k}$ and target space
$T^4$.
For simplicity, we set the radius of the circle equal to one (unit circle).
They are described by a strongly coupled CFT$_2$ whose central charge is
\be\label{eqcc} c = 6\mathtt{k} \sim \frac{1}{G} \gg 1\ee
where $G$ is the three-dimensional Newton's constant.
At finite temperature,  the thermal CFT$_2$ has entropy
\be\label{eqScft} S \sim \mathtt{k} \sim \frac{1}{G} \ee
On the gravity side, the finite temperature is provided by a BTZ black hole~\cite{bibk}.
If the radius of its horizon is $o(1)$, then so is the area of the horizon
($A\sim 1$) and the Bekenstein-Hawking entropy is
\be S = \frac{A}{4G} \sim \frac{1}{G} \ee
in agreement with the CFT result~(\ref{eqScft}).
For such a system, the Poincar\'e recurrence time is estimated to be~\cite{bibq13a}
\be\label{eqex} t_P \sim e^S \sim o(e^{1/G}) \ee
It is clear from this expression that in order to understand the gravity side
of the AdS/CFT correspondence at finite $G\sim 1/\mathtt{k}$, one ought to
include contributions to gravity correlators beyond the semi-classical
approximation which will modify the BTZ black-hole background. A number of alternatives
have been entertained~\cite{bibq13a,bibpr1,bibpr2,bibpr3,bibpr4,bibpr5,bibsol}.

Here we follow a proposal by Solodukhin~\cite{bibsol} and replace the
black hole by a wormhole, thus eliminating the horizon and the attendant leakage of information.
The size of the narrow throat will be
\be\label{eqpro} \lambda \sim o(e^{-1/G}) \ee
leading to a Poincar\'e recurrence time
\be\label{eqtp} t_P \sim \frac{1}{\lambda} \sim o(e^{1/G})\ee
in agreement with expectations (eq.~(\ref{eqex})).
We shall calculate two-point functions explicitly and obtain the {\em real}
poles of the propagator, thus demonstrating unitarity.


We start by reviewing known exact results for the BTZ black hole~\cite{bibq7,bibq12,bibq13}.
The metric for a non-rotating BTZ black hole reads
\be\label{eqmbtz}
ds^2 = -\sinh^2 y \ dt^2 +dy^2+ \cosh^2 y\ d\phi^2
\ee
where $\phi\in [0,2\pi)$. The horizon is at $y=0$ and $T_H = \frac{1}{2\pi}$ is the Hawking temperature.
The wave equation for a massive scalar of mass $m$ is
\be\label{bhweq}
\frac{1}{\sinh y\cosh y}\ \frac{\p}{\p y} \left( \sinh y\cosh y \frac{\p\Phi}{\p y} \right) -\frac{1}{\sinh^2 y}\ \frac{\p^2 \Phi}{\p t^2} + \frac{1}{\cosh^2 y}
\ \frac{\p^2 \Phi}{\p \phi^2} = m^2\Phi
\ee
to be solved outside the horizon ($y>0$).
The solution may be written as
\be
\Phi = e^{i(\omega t-k\phi) }\Psi (y) ,\ \ \ \ \ k \in\mathbb{Z}
\ee
where $\Psi$ satisfies
\be\label{eq5}
\frac{1}{\sinh y\cosh y}\ \left( \sinh y\cosh y \ \Psi' \right)'
+ \left( \frac{\omega^2}{\sinh^2 y} +\frac{k^2}{\cosh^2 y} \right)\Psi =m^2 \Psi
\ee
Two independent solutions are obtained by examining the behavior near the horizon
($y\to 0$),
\be\label{eq11} \Psi_\pm \sim y^{\pm i\omega}\ee
They can be written in closed form in terms of hypergeometric functions,
\be\label{eqpsip} \Psi_\pm = \cosh^{-2h_+} y\ \tanh^{\pm i\omega} y\
F(h_+ \pm\halfi (\omega + k), h_+ \pm\halfi (\omega - k); 1\pm i\omega; \tanh^2 y)\ee
where $h_\pm = \half (1\pm \sqrt{1+m^2})$.
The acceptable solution is $\Psi_-$ (purely in-going at the horizon).
Near the boundary ($y\to \infty$), it behaves as
\be\label{eq11b} \Psi_- \sim \mathcal{A}_+ \ e^{-2h_-y}
+\mathcal{A}_- \ e^{-2h_+y} \ \ , \ \ \mathcal{A}_\pm =
\frac{\Gamma(2h_\pm -1)\Gamma(1- i\omega)}{\Gamma(h_\pm - \halfi (\omega+k))
\Gamma(h_\pm - \halfi (\omega -k))}\ee
For quasi-normal modes, we demand that $\Psi_-$ vanish at the boundary
and therefore set $\mathcal{A}_+ = 0$.
This leads to a discrete spectrum of complex frequencies,
\be\label{eqwme2} \omega_n = \pm k  -i(n +h_+)\quad,\quad n=0,1,2,\dots
\ee
with negative imaginary part, as expected~\cite{bibq2}.
It is perhaps worth mentioning that these frequencies may also be obtained by
a simple monodromy argument which makes use of the {\em unphysical} black-hole
singularity~\cite{bibus2}.

Turning to the AdS/CFT correspondence, the flux at the boundary ($y\to\infty$)
is related to the retarded propagator of the corresponding CFT living on
the boundary. A standard calculation yields
\be \tilde G_R (\omega, k) \sim \lim_{y\to\infty} \frac{F'(y)}{F(y)}\ee
Explicitly,
\bea\label{eq20}
\tilde G_R (\omega, k) &\sim& \frac{\mathcal{A}_-}{\mathcal{A}_+}\nonumber\\
&\sim& \frac{\Gamma (h_+ - \halfi(\omega - k))\Gamma (h_+ - \halfi(\omega + k))}{\Gamma (h_- - \halfi(\omega - k))\Gamma (h_- - \halfi(\omega + k))}\nonumber\\
&\sim& |\Gamma (h_+ - \halfi(\omega - k))\Gamma (h_+ - \halfi(\omega + k))|^2 \nonumber\\
& &\times \sin\pi(h_+ - \halfi(\omega - k))
\sin\pi(h_+ - \halfi(\omega + k))\eea
Plainly, the quasi-normal modes (zeroes of $\mathcal{A}_+$) are poles of the
retarded propagator (since $\tilde G_R\sim 1/\mathcal{A}_+$).
Its fourier transform has an exponentially decaying tail
\be G_R(t) = \int d\omega\ e^{-i\omega t} \tilde G_R \sim e^{-h_+ t}\ee
as $t\to\infty$, exhibiting no Poincar\'e recurrences.
It was pointed out~\cite{bibq13a} that this does not contradict the unitarity of the corresponding CFT, because
the latter effectively lives in infinite space.
This is because a string with winding number $\mathtt{k}$ sees a space of
length
\be\label{eqLeff} L_{eff} \sim \mathtt{k}\ee
The AdS/CFT correspondence for a BTZ black hole only works in the infinite-$\mathtt{k}$ limit.
To consider the large but finite $\mathtt{k}$ case, we turn to a discussion of
the AdS/CFT correspondence for a wormhole.


The wormhole metric is~\cite{bibsol}
\be\label{eqmbtzw}
ds^2 = -(\sinh^2 y + \lambda^2)\ dt^2 + dy^2+ \cosh^2 y\ d\phi^2
\ee
In the limit $\lambda\to 0$, it reduces to the BTZ black hole metric~(\ref{eqmbtz}).
There is no horizon at $y=0$; instead the wormhole has a very narrow throat ($o(\lambda)$) joining two regions of space-time with two distinct boundaries (at
$y\to\pm\infty$, respectively).
Information may flow in both directions through the throat.
This modification is significant near the ``horizon" point $y=0$.
As we approach $y\to 0$, the time-like distance is $ds^2 \approx - \lambda^2 dt^2$,
showing that the time scale of the system is $\sim 1/\lambda$.
This is the order of magnitude of the Poincar\'e recurrence time, i.e., we
expect
\be t_P \sim o(1/\lambda)\ee
as advertised~(eq.~(\ref{eqtp})). The value of $\lambda$ will be fixed upon
comparison with the dual CFT.

The radial wave equation for a massive scalar of mass $m$ is now
\be\label{eq26}
\frac{1}{\cosh y\ (\sinh^2 y+\lambda^2)^{1/2}} \left( \cosh y\ (\sinh^2 y+\lambda^2)^{1/2}\
\Psi'\right)' +\left( \frac{\omega^2}{\sinh^2 y +\lambda^2} + \frac{k^2}{\cosh^2 y}\right)
\ \Psi = m^2\Psi
\ee
to be solved along the entire real axis ($y\in\mathbb{R}$), unlike in the black hole case~(eq.~(\ref{bhweq})), where $y$ was restricted to positive values, since the horizon was
at $y=0$.
We wish to solve this equation in the small-$\lambda$ limit ($\lambda \ll 1$).
To this end, we consider three regions,
\begin{itemize}
\item[$(I)$] $y\gg\lambda$ which includes one of the boundaries,
\item[$(II)$] $y\ll -\lambda$ which includes the other boundary, and
\item[$(III)$] $|y|\ll 1$.
\end{itemize}
We shall solve the wave equation in each region and then match the solutions
in the overlapping regions $\lambda \ll y \ll 1$ and $-1\ll y\ll -\lambda$.
Let us start with region $(II)$.
In this region, the wave equation may be approximated by the corresponding
equation for the BTZ black hole~(\ref{eq5}).
Two independent solutions are then given by~(\ref{eqpsip}). However, in our
case there is no physical requirement dictating a choice based on the small-$y$
behavior, because there is no horizon at $y=0$.
In fact, we ought to chose a linear combination which behaves nicely at the
boundary ($y\to -\infty$). Thus, the acceptable solution in this region is
\be\label{eqPsiII} \Psi_{II} = \cosh^{-2h_+} y\ \tanh^{-i\omega} y\
F(h_+ -\halfi (\omega+k), h_+- \halfi (\omega -k); 2h_+; 1/\cosh^2 y)\ee
It vanishes at the boundary ($\Psi_{II}\sim e^{2h_+y}$ as $y\to -\infty$).
At small $y$, it behaves as
\be\label{eq28} \Psi_{II} \sim \mathcal{B}_+ y^{-i\omega} + \mathcal{B}_- y^{+i\omega}
\ \ , \ \ \mathcal{B}_\pm = \frac{\Gamma (2h_+) \Gamma (\pm i\omega)}{\Gamma(h_+ \pm \halfi (\omega+k))\Gamma(h_+ \pm \halfi (\omega -k))}\ee
In region $(III)$ ($|y|\ll 1$), the wave equation~(\ref{eq26}) reduces to
\be\label{eq26a}
\frac{1}{(y^2+\lambda^2)^{1/2}} \left( (y^2 +\lambda^2)^{1/2}\
\Psi_{III}'\right)' + \frac{\omega^2}{y^2 + \lambda^2}
\ \Psi_{III} = 0
\ee
Two linearly independent solutions are
\be \Psi_{III}^{(1)} = F(\halfi\omega, -\halfi\omega ; \half ; -y^2/\lambda^2)\ \ , \ \
\Psi_{III}^{(2)} = \frac{y}{\lambda}\, F(\half + \halfi\omega, \half -\halfi\omega ; \three ; -y^2/\lambda^2) \ee
At large $y>0$, they behave as
\be\label{eq31} \Psi_{III}^{(1)} \sim \frac{1}{2} \left( \frac{2y}{\lambda} \right)^{+i\omega} + \frac{1}{2} \left( \frac{2y}{\lambda} \right)^{-i\omega}
\ \ , \ \
\Psi_{III}^{(2)} \sim \frac{i}{2\omega} \left( \frac{2y}{\lambda} \right)^{+i\omega}
 - \frac{i}{2\omega} \left( \frac{2y}{\lambda} \right)^{-i\omega}
\ee
Analytically continuing to negative $y$, we note that $\Psi_{III}^{(1)}$ is an
even function whereas $\Psi_{III}^{(2)}$ is odd.
We therefore obtain the asymptotic behavior
\be \Psi_{III}^{(1)} \sim \frac{1}{2} \left( \frac{2y}{\lambda} \right)^{+i\omega} + \frac{1}{2} \left( \frac{2y}{\lambda} \right)^{-i\omega}
\ \ , \ \
\Psi_{III}^{(2)} \sim -\frac{i}{2\omega} \left( \frac{2y}{\lambda} \right)^{+i\omega}
 + \frac{i}{2\omega} \left( \frac{2y}{\lambda} \right)^{-i\omega}
\ee
as $y\to -\infty$.
Matching this to the asymptotic behavior of $\Psi_{II}$ given by eq.~(\ref{eq28}), we obtain the acceptable solution in region $(III)$,
\be \Psi_{III} = \mathcal{B}_+ \Psi_{III}^{(-)} + \mathcal{B}_- \Psi_{III}^{(+)}
\ \ , \ \ \Psi_{III}^{(\pm)} = \left( \frac{2}{\lambda} \right)^{\mp i\omega}
\left( \Psi_{III}^{(1)} \pm i\omega \Psi_{III}^{(2)} \right)\ee
On account of~(\ref{eq31}), at large $y>0$, it behaves as ({\em cf.}~eq.~(\ref{eq28}))
\be\label{eq31a}
\Psi_{III} \sim \mathcal{B}_+ \left( \frac{4y}{\lambda^2} \right)^{+i\omega} + B_- \left( \frac{4y}{\lambda^2} \right)^{-i\omega}
\ee
Finally, in region $(I)$, two linearly independent solutions are ({\em cf.}~eq.~(\ref{eqPsiII}) in region $(II)$)
\be\label{eqPsiI} \Psi_{I}^{(\pm )} = \cosh^{-2h_\pm } y\ \tanh^{-i\omega} y\
F(h_\pm  -\halfi (\omega+k), h_\pm - \halfi (\omega -k); 2h_\pm ; 1/\cosh^2 y)\ee
Matching the asymptotic behavior~(\ref{eq31a}), we obtain the solution in
region $(I)$,
\be \Psi_{I} = \alpha_+ \Psi_{I}^{(-)} + \alpha_- \Psi_{I}^{(+)} \ee
where
\be \alpha_+ = \frac{\mathcal{B}_+^2 \left( \frac{2}{\lambda} \right)^{2i\omega}
- \mathcal{B}_-^2 \left( \frac{2}{\lambda} \right)^{-2i\omega}}{\mathcal{B}_+
\mathcal{C}_- - \mathcal{B}_- \mathcal{C}_+}\ \ , \ \
\alpha_- = \frac{\mathcal{B}_- \mathcal{C}_- \left( \frac{2}{\lambda} \right)^{-2i\omega}- \mathcal{B}_+ \mathcal{C}_+ \left( \frac{2}{\lambda} \right)^{2i\omega}}{\mathcal{B}_+
\mathcal{C}_- - \mathcal{B}_- \mathcal{C}_+}\ee
$\mathcal{B}_\pm$ are given by (\ref{eq28}) and
\be\label{eq28c}
\mathcal{C}_\pm = \frac{\Gamma (2h_-) \Gamma (\pm i\omega)}{\Gamma(h_- \pm \halfi (\omega+k))\Gamma(h_- \pm \halfi (\omega -k))}\ee
For a normalizable solution, we ought to set $\alpha_+ = 0$, which leads to
the quantization condition
\be\label{eqpol} \left( \frac{2}{\lambda} \right)^{2i\omega} = \frac{\mathcal{B}_-}{\mathcal{B}_+} =
\frac{\Gamma (- i\omega)\Gamma(h_+ + \halfi (\omega +k))\Gamma(h_+ + \halfi (\omega -k))}{\Gamma (+ i\omega)\Gamma(h_+ - \halfi (\omega +k))\Gamma(h_+ - \halfi (\omega -k))}\ee
This leads to a discrete spectrum of {\em real} frequencies (notice that both
sides of the equation have unit norm, since $\mathcal{B}_-$ is the complex
conjugate of $\mathcal{B}_+$ for real $\omega$).
For small $\omega$, we may approximate $\mathcal{B}_-/\mathcal{B}_+
\approx -1$, therefore we obtain the spectrum
\be\label{eq35} \omega_n \approx \left( n+ \half \right) \frac{\pi}{\ln \frac{2}{\lambda}}\ \ , \ \ n\in \mathbb{Z}\ee
corresponding to periodicity with period $L_{eff} \sim \ln (1/\lambda)$~\cite{bibq13a}.
Comparing with the CFT result that the periodicity is $L_{eff} \sim \mathtt{k} \sim 1/G$ (eq.~(\ref{eqLeff})),
we deduce
\be \lambda \sim  o( e^{-1/G} ) \ee
as promised (eq.~(\ref{eqpro})),
showing that the wormhole is produced by non-perturbative quantum effects.
Notice also that the period of the CFT propagator $L_{eff}$ is much smaller than
the Poincar\'e recurrence time (eq.~(\ref{eqex})),
\be L_{eff} \ll t_P \ee
This is reminiscent of the chaotic behavior of the brick wall modification
of the BTZ black hole~\cite{bibsol} and is attributed to the rich structure of the spectrum
given by~(\ref{eqpol}).

The retarded propagator is
\be \tilde G_R (\omega,k) \sim \frac{\alpha_-}{\alpha_+} \sim
\frac{\mathcal{B}_- \mathcal{C}_- - \mathcal{B}_+ \mathcal{C}_+ \left(
\frac{2}{\lambda} \right)^{4i\omega}}{\mathcal{B}_-^2 - \mathcal{B}_+^2 \left(
\frac{2}{\lambda} \right)^{4i\omega}}
\ee
and has real poles given by~(\ref{eqpol}).
It is defined in the upper-half complex $\omega$-plane. In the limit $\lambda\to 0$
(or, equivalently, $\mathtt{k} \to\infty$), the spectrum of real frequencies~(\ref{eq35})
becomes continuous, signaling the emergence of a horizon. The retarded propagator becomes
\be \tilde G_R (\omega,k) \sim \frac{\mathcal{C}_-}{\mathcal{B}_-}
\ee
which agrees with our earlier BTZ result (eq.~(\ref{eq20})), since $\mathcal{B}_- \sim \mathcal{A}_+$
and $\mathcal{C}_- \sim \mathcal{A}_-$
(using eqs.~(\ref{eq11b}), (\ref{eq28}) and (\ref{eq28c})).
In this limit, the quasi-normal modes~(\ref{eqwme2}) emerge as poles (zeroes of
$\mathcal{B}_- \sim \mathcal{A}_+$).
It should be emphasized that for no other value of $\lambda$, no matter how
small, do complex poles arise.


To summarize, we replaced the BTZ black hole with a wormhole~\cite{bibsol} in order to
study the AdS$_3$/CFT$_2$ correspondence at finite temperature and large but
finite central charge~(\ref{eqcc}) of the CFT.
The wormhole had a narrow throat of size $\lambda$. We argued that the system,
once perturbed, exhibited Poincar\'e recurrences with time constant $t_P\sim
1/\lambda$. We solved the wave equation for a scalar field in the small-$\lambda$ limit
and deduced the propagator for the dual CFT. We found an explicit equation for
the poles~(eq.~(\ref{eqpol})) which yielded a rich spectrum of {\em real} eigenvalues
demonstrating the unitarity of time evolution. Upon comparison with the
expected periodicity of the dual CFT, we deduced that the wormhole parameter
$\lambda$ was of order $e^{-1/G}$, suggesting that the wormhole can only emerge
through non-perturbative effects.

%

\end{document}